\documentstyle[graphics]{caosp}
\begin{document}
\pubyear{1998}
\volume{27}
\firstpage{390}
\htitle{Supercomputing and Ap stars}

\hauthor{Martin~J.~Stift}

\title{Supercomputing and Ap stars}

\author{Martin~J.~Stift}

\institute{Institut f{\"u}r Astronomie, T{\"u}rkenschanzstr. 17,
A-1180 Wien, Austria}


\maketitle

\begin{abstract}
Certain problems in the field of stellar atmospheres, polarised 
radiative transfer and magnetic field diagnostics cannot be 
addressed by means of traditional sequential programming techniques 
because CPU times become prohibitive on even the fastest single 
processor machines when realistic physics and accurate numerical methods
are employed. This contribution discusses the question of what kind of
supercomputing approach is best suited for the modelling of Ap stars, 
pointing out the superiority of parallel computing with Ada95 over 
High Performance Fortran in all of the above-mentioned fields.

\keywords{Stars: magnetic fields -- Polarization -- Methods: numerical}

\end{abstract}

\section{Introduction}
\label{intr}

In recent times, problems in the modelling of both nonmagnetic and 
magnetic stellar atmospheres have emerged that cannot be solved by 
traditional numerically intensive computing. Take as an example the 
calculation of line-blanketed LTE stellar atmospheres by means of 
opacity sampling as done in {\sl ATLAS12}. Even on present-day fast 
single-processor machines realistic frequency step sizes lead to rather 
prohibitive CPU times (see Castelli, these proceedings). The same 
applies to the modelling of broadband linear (BBLP) and circular (BBCP) 
polarisation in sunspots, in the solar network and in magnetic stars, 
and to full Zeeman Doppler Imaging (ZDI). The usual remedies to this 
software crisis do not excel in imagination: employing coarse frequency 
grids, approximate formal solvers or Milne-Eddington atmospheres is a 
cheap expedient but does not attack the problem at its root.

Instead of waiting for the next increase in CPU clock rates to be able
to run today's models on tomorrow's computers, it would be preferable 
to go parallel. Indeed, parallel architectures are readily available 
nowadays and languages with parallel constructs provide concurrent 
execution of program segments. Obviously there is no way around the 
restructuring and at least partial rewriting of existing programs, but
isn't this preferable to physically doubtful approximations or to poor
frequency sampling?

High Performance Fortran (HPF) would seem the obvious choice for the
denizens of the Fortran universe, but does the data parallel paradigm 
of HPF really provide optimum parallelism in spectral line synthesis, 
be it LTE or NLTE, polarised or unpolarised? Would HPF really speed up
stellar atmosphere calculations with a suitably modified version of 
{\sl ATLAS}, and what about ZDI? Isn't it most likely that we have to 
rethink our approach in a far more radical way?

\section{Not just number crunching}
\label{Notjust}

Supercomputing should not be a synonym for brute number crunching 
nor should it reduce to the use of a few special instructions and
highly specialised subprograms that can take advantage of parallel 
architectures. Supercomputing deserves its name only when it 
encompasses object oriented software design on the {\em appropriate 
level of abstraction}, when it ensures code safety and reliability, 
when it provides potentially massively parallel execution of 
{\em large sections} of the code, and when the most accurate and 
stable numerical methods are used.

\begin{figure}[h]
\begin{center}
\scalebox{0.6}[0.4]{                                        
\includegraphics*[0.8cm,5.5cm][20.0cm,15.0cm]{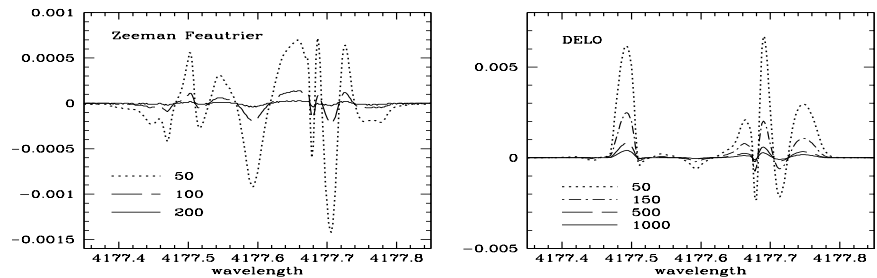}}
\end{center}
\caption{Differential linear polarisation spectra (Stokes $Q$)
calculated for a solar atmospheric model, a field strength of 
4\,T and an angle between field vector and line of sight of 
$45\degr$. All results -- ZF based (left) or DELO based (right)
-- are displayed relative to the reference result (ZF with 500 
depth points) and labeled with the number of depth points.  
{\em Note the difference in the scales (factor~5)!}}
\end{figure}                                           

In polarised radiative transfer the latter requirement translates
into the use of the Zeeman Feautrier (ZF) method (Auer et al. 1977). 
Tedious to code and prone to bugs in Fortran77, a ZF solver 
constitutes no problem for the Ada programmer thanks to the high 
level abstractions made possible by the use of the Ada programming 
language; the block tri-diagonal scheme can be written down 
straightforwardly as given in Rees \& Murphy (1987). Extensive 
tests have revealed that especially in the presence of blends 
the ZF solver is (at constant number of depth points) up to 
5~times more accurate than the DELO method (Rees \& Murphy, 1987) 
as demonstrated in Fig.~1. If a 4000\,{\AA} interval is to be covered,
the modelling of BBLP and BBCP in a solar-type atmosphere involves 
opacity sampling over about $4.5\,10^5$ Zeeman subcomponents and 
$4\,10^5$ formal solutions to achieve the minimum frequency resolution.
Since it is well known that in the presence of heavy blending 
Milne-Eddington based approaches are hopelessly inadequate we have 
no alternative to the admittedly expensive ZF and DELO solvers. The 
only way to be able to afford those relatively slow solvers appears 
to lie in parallelism on a large scale.

\section{Questions of technology : parallelism}
\label{Tech}

You don't have to visit homepages of astronomical colleagues to know 
that most of them program in Fortran. But is Fortran (or HPF) of any 
help in the computational astrophysics problems listed above? Codes 
of some 1000 Fortran statements have been written ab initio in the 
late 1980s using simplified formal solvers and coarse fixed spatial 
integration grids to synthesise intensity spectra over intervals a 
mere 2\,{\AA} wide; compare this to the hundreds and thousands of 
{\AA}ngstr{\"o}ms required for the modelling of broadband
polarisation. Can such a program be upgraded for the latter purpose?

Obviously a minor change won't do it but even if one were prepared
to restructure large parts of the program, I claim that for very 
fundamental reasons this is not possible in a purely Fortran context.
There are no threads of control in {\em data parallel} HPF, so there
is no way to directly implement the natural approach, viz. the 
computation in parallel of the emerging spectrum at each frequency 
point. For this one would have to employ POSIX threads (pthreads),
taking care of the individual threads, mutexes, and locks, but this
is truly hard and unrewarding work. None of the HPF features promise 
substantial gains in performance.

Ada95 and its concurrent constructs, the {\sf task} types and the 
{\sf protected} types are ideally suited for parallelising line 
synthesis and stellar atmosphere codes. Task objects are program 
entities that can execute concurrently on different nodes (also in 
distributed systems), protected objects can be used to provide 
light-weight synchronisation. 

\section{Ada95: object orientation and parallelism at work}
\label{Work}

In the past few years I have developed a new generation of codes 
in the fields of polarised and unpolarised line synthesis and of 
Zeeman Doppler Imaging. These codes are all written in Ada95 and 
conform to my definition of supercomputing given above. They 
incorporate up-to-date astrophysics, deal with realistic atmospheres, 
provide full treatment of blends involving anomalous Zeeman patterns, 
and offer a choice of accurate and numerically stable formal solvers 
(DELO, ZF). On the software side, maximum reuse of software modules 
is achieved by information hiding and encapsulation, and by extensive 
use of generics, of child libraries and of inheritance. And finally,
all codes provide for potentially massive parallelism; they run 
-- with virtually no change -- on anything from PCs to Silicon Graphics
supercomputers, taking full advantage of resources ranging from 
1~processor to 32~processors and more.

\begin{figure}[tbh]
\begin{center}
\scalebox{0.57}[0.65]{                                        
\includegraphics*[0.8cm,16.0cm][20.0cm,21.0cm]{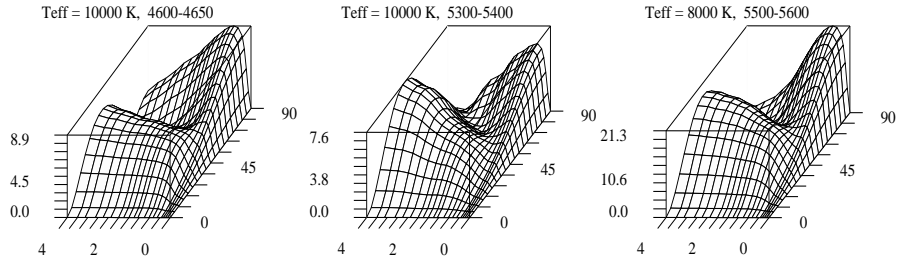}}
\caption{Degree of broadband linear polarisation $p$ as a function
of magnetic field strength (from 0 to 4~Tesla) and angle between 
field vector and line of sight (from 0 to 90~degrees) for different 
model atmospheres and wavelength intervals.}
\end{center}
\end{figure}                                           

It is absolutely amazing how congenial the {\em control-parallel} 
paradigm of Ada tasking is to line synthesis and stellar atmosphere 
problems. No large-scale modifications are needed to convert the 
sequential version of a program to a parallel version: changing
not even 20 expressions in a 3500~LOC (lines of code) Ada program 
is sufficient to obtain a simple parallel version of a sequential 
code. At the same time, almost perfect load balance (distribution 
of the computations to the various CPUs according to their 
availability) is achieved in an easy and elegant way through 
the use of protected objects for synchronisation. In most cases
there is nothing more to do than put the subprogram that is to
be executed in parallel into a task, replace those variables that
are to be read or updated in mutual exclusion by protected objects,
and finally statically create or dynamically allocate as many
task objects as processors are available. In the shortest of times
(2 hours and less) you are gratified with a parallel program!
Numerous examples of how the object oriented and parallel features
of Ada95 can be employed in scientific computing can be found in 
Stift (1998).

\section{Supercomputing results on CP stars and outlook}
\label{Results}

Thanks to a dedicated Silicon Graphics Origin200 server with
four R10000 processors, magnetic broadband polarisation in heavily
blended spectra is at last revealing some of its secrets.
Calculations of extensive grids of broadband polarisation as
a function of magnetic field strength and direction, of the
atmospheric model, and of the wavelength interval have shown that 
the polarisation signal does not saturate at large field values 
as in the case of the classical Zeeman triplet but may display 
complex behaviour as can be seen in Fig.~2. It thus appears that 
strong fields do not necessarily lead to a strong polarisation 
signal, but that rather the opposite can be true. 

Leroy (1989) has demonstrated that the wavelength dependence of 
the degree of linear polarisation $p$ depends on the magnetic field 
strength; his analysis did not include the effects of blending.
Extending the calculations presented by Stift (1997) by synthesising 
spectra over the whole visible range for various atmospheres and a 
grid of magnetic field strengths and field directions (Fig.~3 
displays the solar case) allows a systematic investigation of the 
$p$ vs. $\lambda$ relation. As Franco Leone has suggested, this 
relation could possibly be used as a diagnostic tool for estimating 
the mean magnetic field modulus of a magnetic Ap star. First results 
indicate that this is indeed the case, the $p - \lambda$ relation 
appearing to be insensitive to the stellar magnetic geometry and 
independent of magnetic phase, reflecting only the mean magnetic 
field modulus.

\begin{figure}[tbh]
\begin{center}
\scalebox{0.63}[0.52]{                                        
\includegraphics*[2.0cm,5.9cm][20.5cm,13.3cm]{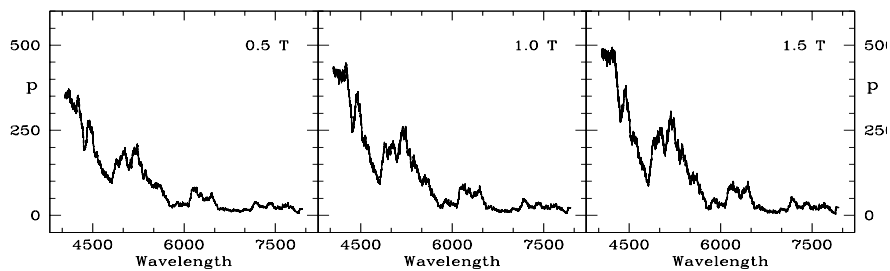}}
\end{center}
\caption{The relation between degree of linear polarisation $p$ and
wavelength $\lambda$ as a function of magnetic field strength $B$
in the purely transversal case and a solar atmosphere. The
polarisation signal has been integrated over 100{\AA} intervals
and smoothed with a 400 pixel box function for clarity's sake.}
\end{figure}                                           

The outlook is fascinating: supercomputing with Ada95 provides the
means for major advances in the field of magnetic polarisation and
stellar atmospheres, combining object orientation with
straightforward scalable parallelism. With the technology and
thousands of lines of Ada code available for free, computational
astrophysics can easily overcome the present software crisis.

\acknowledgements
This work has been made possible by the Austrian {\it Fonds zur 
F{\"o}rderung der wissenschaftlichen Forschung} under project
P12101-AST. Additional support came from the 
{\it Hochschuljubil{\"a}umsstiftung der Stadt Wien}.

\end{document}